\documentstyle[12pt]{article}
\catcode`\@=11
\long\def\@makefntext#1{
\protect\noindent \hbox to 3.2pt {\hskip-.9pt
$^{{\ninerm\@thefnmark}}$\hfil}#1\hfill}		

 \def\@makefnmark{\hbox to 0pt{$^{\@thefnmark}$\hss}}  

\def\ps@myheadings{\let\@mkboth\@gobbletwo
\def\@oddhead{\hbox{}
\rightmark\hfil\ninerm\thepage}
\def\@oddfoot{}\def\@evenhead{\ninerm\thepage\hfil
\leftmark\hbox{}}\def\@evenfoot{}
\def\sectionmark##1{}\def\subsectionmark##1{}}

\newcounter{sectionc}\newcounter{subsectionc}\newcounter{subsubsectionc}
\renewcommand{\section}[1] {\vspace{0.6cm}\addtocounter{sectionc}{1}
\setcounter{subsectionc}{0}\setcounter{subsubsectionc}{0}\noindent
	{\bf\thesectionc. #1}\par\vspace{0.4cm}}
\renewcommand{\subsection}[1] {\vspace{0.6cm}\addtocounter{subsectionc}{1}
	\setcounter{subsubsectionc}{0}\noindent
	{\it\thesectionc.\thesubsectionc. #1}\par\vspace{0.4cm}}
\renewcommand{\subsubsection}[1]
{\vspace{0.6cm}\addtocounter{subsubsectionc}{1}
	\noindent {\rm\thesectionc.\thesubsectionc.\thesubsubsectionc.
	#1}\par\vspace{0.4cm}}

\newcounter{appendixc}
\newcounter{subappendixc}[appendixc]
\newcounter{subsubappendixc}[subappendixc]

\renewcommand{\appendix}[1] {\vspace{0.6cm}
        \refstepcounter{appendixc}
        \setcounter{figure}{0}
        \setcounter{table}{0}
        \setcounter{equation}{0}
        \renewcommand{\thefigure}{\Alph{appendixc}.\arabic{figure}}
        \renewcommand{\thetable}{\Alph{appendixc}.\arabic{table}}
        \renewcommand{\theappendixc}{\Alph{appendixc}}
        \renewcommand{\theequation}{\Alph{appendixc}.\arabic{equation}}
        \noindent{\bf Appendix \theappendixc #1}\par\vspace{0.4cm}}

\def\abstracts#1{{
	\centering{\begin{minipage}{30pc}\tenrm\baselineskip=12pt\noindent
	\centerline{\tenrm ABSTRACT}\vspace{0.3cm}
	\parindent=0pt #1
	\end{minipage}}\par}}

\renewenvironment{thebibliography}[1]
	{\begin{list}{\arabic{enumi}.}
	{\usecounter{enumi}\setlength{\parsep}{0pt}
\setlength{\leftmargin 1.25cm}{\rightmargin 0pt}
	 \setlength{\itemsep}{0pt} \settowidth
	{\labelwidth}{#1.}\sloppy}}{\end{list}}

\newcounter{itemlistc}
\newcounter{romanlistc}
\newcounter{alphlistc}
\newcounter{arabiclistc}

\newcommand{\fcaption}[1]{
        \refstepcounter{figure}
        \setbox\@tempboxa = \hbox{\tenrm Fig.~\thefigure. #1}
        \ifdim \wd\@tempboxa > 6in
           {\begin{center}
        \parbox{6in}{\tenrm\baselineskip=12pt Fig.~\thefigure. #1}
            \end{center}}
        \else
             {\begin{center}
             {\tenrm Fig.~\thefigure. #1}
              \end{center}}
        \fi}

\newcommand{\tcaption}[1]{
        \refstepcounter{table}
        \setbox\@tempboxa = \hbox{\tenrm Table~\thetable. #1}
        \ifdim \wd\@tempboxa > 6in
           {\begin{center}
        \parbox{6in}{\tenrm\baselineskip=12pt Table~\thetable. #1}
            \end{center}}
        \else
             {\begin{center}
             {\tenrm Table~\thetable. #1}
              \end{center}}
        \fi}

\def\@citex[#1]#2{\if@filesw\immediate\write\@auxout
	{\string\citation{#2}}\fi
\def\@citea{}\@cite{\@for\@citeb:=#2\do
	{\@citea\def\@citea{,}\@ifundefined
	{b@\@citeb}{{\bf ?}\@warning
	{Citation `\@citeb' on page \thepage \space undefined}}
	{\csname b@\@citeb\endcsname}}}{#1}}

\newif\if@cghi
\def\cite{\@cghitrue\@ifnextchar [{\@tempswatrue
	\@citex}{\@tempswafalse\@citex[]}}
\def\citelow{\@cghifalse\@ifnextchar [{\@tempswatrue
	\@citex}{\@tempswafalse\@citex[]}}
\def\@cite#1#2{{$\null^{#1}$\if@tempswa\typeout
	{IJCGA warning: optional citation argument
	ignored: `#2'} \fi}}

\def\fnt#1#2{\footnotetext{\kern-.3em
	{$^{\mbox{\sevenrm #1}}$}{#2}}}

 1
 1
 1

\font\tenbf=cmbx10
\font\tenrm=cmr10
\font\tenit=cmti10

\font\ninerm=cmr9

\newcommand{\bea}{\begin{eqnarray}}
\newcommand{\eea}{\end{eqnarray}}
\newcommand{\beq}{\begin{equation}}
\newcommand{\eeq}{\end{equation}}
\newcommand{\bec}{\begin{center}}
\newcommand{\eec}{\end{center}}
\newcommand{\beqn}{\begin{displaymath}}
\newcommand{\eeqn}{\end{displaymath}}
\newcommand{\zeilea}{\renewcommand{\baselinestretch}{1.5}
                     \small\normalsize}
\newcommand{\zeilee}{\renewcommand{\baselinestretch}{1}
                     \small\normalsize}
\newcommand{\vslash}{\mbox{$\not{\hspace{-0.7mm}v}$}}           

\newcommand{\kslash}{\mbox{$\not{\hspace{-1mm}k}$}}             
\newcommand{\eslash}{\mbox{$\not{\hspace{-0.8mm} \varepsilon}$}}

\newcommand{\eins}{\mbox{$\rule{2.5mm}{0.1mm}
                          {\hspace{-2.7mm}1}
                          {\hspace{-0.2mm}\rule{0.07mm}{2.7mm}}$}}

\newcommand{\mini}{\mbox{\scriptsize min}}
\newcommand{\Yk}  {\begin{Young}
                      $\scriptstyle{k}$\cr
                      \end{Young} }
\newcommand{\YK}  {\begin{Young}
                      $\scriptscriptstyle{K}$\cr
                      \end{Young} }

\newcommand{\Yzehn}{\begin{Young}
                      &\cr
		      \end{Young} }
\newcommand{\Yeins}  {\begin{Young}
		      \cr
                      \cr
                      \cr
                      \end{Young} }
\newcommand{\Yanti}    {\begin{Young}
                      \cr
                      \cr
                      \end{Young} }
\newdimen\hoogte    \hoogte=7pt    
\newdimen\breedte   \breedte=9pt   
\newdimen\dikte     \dikte=0.5pt    
\def\beginYoung{
       \begingroup
       \def\vr{\vrule height0.8\hoogte width\dikte depth 0.2\hoogte}
       \def\fbox##1{\vbox{\offinterlineskip
                    \hrule height\dikte
                    \hbox to \breedte{\vr\hfill##1\hfill\vr}
                    \hrule height\dikte}}
       \vbox\bgroup \offinterlineskip \tabskip=-\dikte \lineskip=-\dikte
            \halign\bgroup &\fbox{##\unskip}\unskip  \crcr }
\def\End@Young{\egroup\egroup\endgroup}
\newenvironment{Young}{\beginYoung}{\End@Young}
\begin{document}

\baselineskip=22pt
\centerline{\tenbf HEAVY BARYONS: CURRENT, PION AND PHOTON TRANSITIONS}

\vspace{0.8cm}
\centerline{\tenrm J.G. K\"ORNER\footnote[1]
{Supported in part by  the BMFT, FRG, under contract 06MZ730}}
\vspace{0.3cm}
\centerline{\tenrm and}
\vspace{0.3cm}
\centerline{\tenrm J. LANDGRAF}
\vspace{0.5cm}
\baselineskip=13pt
\centerline{\tenit Institut f\"ur Physik, Johannes Gutenberg-Universit\"at}
\baselineskip=13pt
\centerline{\tenit Staudinger Weg 7, D-55099 Mainz, Germany}
\vspace{0.9cm}
\abstracts{We discuss the structure of current-induced bottom baryon to
charm baryon transitions, and the structure of pion and photon transitions
between heavy charm or bottom baryons in the Heavy Quark Symmetry limit as
${\scriptstyle m_Q\rightarrow \infty}$. Our discussion involves the ground
state $s$-wave heavy baryons  as well as the excited $p$-wave heavy baryon
states.}

\vfil
\rm\baselineskip=14pt
\section{Introduction}
The Heavy Quark Effective Theory (HQET) formulated in 1990\cite{landa} is
so well known by now that it no longer needs an extensive introduction. The
HQET provides a systematic expansion of QCD in terms of inverse powers of
the heavy quark mass. The leading term in this expansion gives rise to a
new spin  and flavour symmetry at equal velocities, termed Heavy Quark
Symmetry. Corrections to the Heavy Quark Symmetry limit can be classified and
evaluated order by order in $1/m_Q$ by considering the contributions of the
nonleading terms
in the effective HQET fields and the HQET Langrangian. We might mention
that there exist different formulations of HQET which differ from one
another starting at $O(1/m_Q^2)$\cite{landb,landc}. They can be transformed
into each other by appropriate field redefinitions \cite{landd}. Among these,
the Foldy-Wouthuysen type HQET introduced in \cite{landc} is closest to the
original quantum mechanical version of the Foldy-Wouthuysen transformation
introduced in the 50's to deal with recoil corrections when calculating e.g.
properties of the hydrogen atom.

In this review we will only be concerned with the leading order term in the
HQET expansion, i.e. in the Heavy Quark Symmetry limit. In this limit the
dynamics of the light and heavy constituents decouple and the calculation
of the Heavy Quark Symmetry predictions for transition amplitudes
essentially amounts to an angular momentum coupling exercise, however
involved it may be. It is then not surprising that such calculations have
been done before even before the conceptual foundations of HQET had been laid
down in 1990. For example, the Heavy Quark Symmetry structure of
current-induced charm baryon transitions had been written down as early as
1976\cite{lande}. Also, and this will be explicated in this talk, the
technical tools needed to do the requisite angular momentum coupling
calculations already existed for quite some time in the form of the Wigner
6-$j$-symbol calculus developed and used extensively by atomic and nuclear
physicists working on problems closely related to Heavy Quark Symmetry
calculations.

At present most of the attention of experimentalists and theoreticians
working in heavy quark physics is directed towards the application of HQET
in the meson sector where data is starting to become quite abundant. This
data will be supplemented in the not-too-distant future by corresponding
data on heavy baryon decays and there will be a need to analyze this data
in terms of HQET as applied to heavy baryons. This will be the subject of
the present review. We shall discuss the structure of flavour changing
bottom baryon to charm baryon decays, as well as pion and photon transition
between heavy baryons with the same flavour, all in the Heavy Quark
Symmetry limit as $m_Q\rightarrow\infty$. Our discussion involves both
ground state heavy baryons and their $p$-wave excitations. We present our
results in the form of building blocks listed in tabular form. The building
blocks may then easily be assembled to obtain the Heavy Quark Symmetry
predictions for the current, pion and photon transition amplitudes of heavy
baryons.

\section{Classification of $s$- and $p$-Wave Heavy Baryon States}
A heavy baryon is made up of a light diquark system $(qq)$ and a heavy
quark Q. The light diquark system has bosonic quantum numbers $j^P$ with
the total angular momentum $j=0,1,2 \dots $ and parity $P=\pm 1$. To each
diquark system with spin-parity $j^P$ there is a degenerate heavy baryon
doublet with $J^P=(j\pm1/2)^P$ ($j=0$ is an exception). It is important to
realize that the Heavy Quark Symmetry structure of the heavy baryon states
is entirely determinated by the spin-parity $j^P$ of the light diquark
system.

{}From our experience with light baryons and light mesons we know that one
can get a reasonable description of the light particle spectrum in the
constituent quark model picture. This is particularly true for the
enumeration of states, their spins and their parities. As much as we
know up to now, gluon degrees of freedom do not seem to contribute to the
particle spectrum. It is thus quite natural to try the same constituent
approach to enumerate the light diquark states, their spins and their
parities. From the spin degrees of freedom of the two light
quarks one obtains a spin 0 and a spin 1 state. The total orbital state of
the diquark system is characterized by two angular degrees of freedom
which
we take to be the two independent relative momenta $k=\frac{1}{2}(p_{1}-
p_{2})$ and $K=\frac{1}{2}(p_{1}+p_{2}-2p_{3})$ that can be formed from the
two light quark momenta $p_{1}$ and $p_{2}$ and the heavy quark momentum
$p_{3}$. The k-orbital momentum describes relative orbital exitations of
the two quarks, and the K-orbital momentum describes orbital excitations of
the center of mass of the two light quarks relative to the heavy quark.
The $(k,K)$ basis is quite convenient for two reasons. First, Copley, Isgur
and Karl\cite{landf} have found that the $(k,K)$ basis diagonalizes the
Hamiltonian, when harmonic interquark forces are used. Second, it allows one
to classify the diquark states in terms of $SU(2N_f) \otimes O(3)$
representations.

Let us do just this for the two flavour case $N_f=2$ for the $s$- and
$p$-wave states.
\begin{itemize}
\item[a)] $s$-wave gound state $\sim
 \Yzehn_{\,10} \otimes \Yeins_{\,1}$\\
 where the spin-flavour content of the $SU(4)$ diquark representation
 $\underline{10}$ can be determined by looking at how the $\underline{10}$
 representation decomposes under the $SU(2)_{spin} \otimes SU(2)_{flavour}$
 subgroup. One has
 \begin{equation}
 \Yzehn_{\,10}=\left( \;\; \Yanti_{\,1} \otimes \Yanti_{\,1} + \Yzehn_{\,3}
 \otimes \Yzehn_{\,3} \;\right)
 \end{equation}
 When coupling in the heavy quark one finally has the particle content
 \zeilea
 \bea
 \left[ q_1 q_2 \right] : 0^+ &\rightarrow&{\;\,\textstyle \frac{1}{2}^+}
  \quad \quad \! \Lambda_Q  \nonumber\\
 \left\{ q_1 q_2 \right\}:1^+ &\nearrow \atop \searrow&
 \begin{array}{l}
 \frac{1}{2}^+\\
 \frac{3}{2}^+\\
 \end{array}\quad
 \begin{array}{l}
 \Sigma_Q\\
 \Sigma_Q^*\\
 \end{array}
 \eea
 \zeilee
\item[b)]$p$-wave ($l_k=0$, $l_K=1$) $\sim \Yzehn_{\,10} \otimes \YK_{\,3}$:
 The $\underline{10}$ representation decomposes as   before. The
 spin~0 and spin~1 pieces of the $\underline{10}$ couple with $l_K=1$ to
 give $j^P=1^-$ and $j^P=0^-,1^-,2^-$, respectively. One has the particle
 content

 \bea
 \left[ q_1 q_2 \right]:1^- &\nearrow \atop \searrow&
   \left.\begin{array}{l}
   \frac{1}{2}^-\\
   \frac{3}{2}^-\\
   \end{array}
   \right\} \quad \{ \Lambda_{QK1}^{**} \}\nonumber\\
 \left\{ q_1 q_2 \right\}:0^-&\rightarrow& \;\; {\textstyle{\frac{1}{2}^-}}
   \quad\quad\;\;\Sigma_{QK0}^{**} \nonumber\\
 1^-&\nearrow \atop \searrow&
   \left.\begin{array}{l}
   \frac{1}{2}^-\\
   \frac{3}{2}^-\\
   \end{array}\right\}
   \quad \{ \Sigma_{QK1}^{**} \}\nonumber\\
 2^-&\nearrow \atop \searrow&
   \left.\begin{array}{l}
   \frac{3}{2}^-\\
   \frac{5}{2}^-\\
   \end{array}\right\}
   \quad \{\Sigma_{QK2}^{**} \}\;.
 \eea
 \zeilee
\item[c)]$p$-wave ($l_k=1$, $l_K=0$) $\sim \Yanti_{\,6} \otimes \Yk_{\,3}$:
 The diquark representation $\underline 6$ of $SU(4)$ decomposes under the
 $SU(2)_{spin} \otimes SU(2)_{flavour}$ subgroup as
 \begin{equation}
 \Yanti_{\,6}=\left(\;\; \Yanti_{\,1} \otimes \Yzehn_{\,3} + \Yzehn_{\,3}
 \otimes \Yanti_{\,1}\; \right)\;.
 \end{equation}
 After coupling with the $k$-orbital angular momentum $l_k=1$ and the heavy
 quark spin $s=1/2$, the particle content can then be determined to be

 \zeilea
 \bea
 \left[ q_1 q_2 \right]:0^-&\rightarrow& \;\; {\textstyle{\frac{1}{2}^-}}
   \quad \quad \,\; \Lambda_{Qk0}^{**}\nonumber\\
 1^-&\nearrow \atop \searrow&
   \left.\begin{array}{l}
   \frac{1}{2}^-\\
   \frac{3}{2}^-\\
   \end{array}\right\}\quad
   \{ \Lambda_{Qk1}^{**}\} \nonumber\\
 2^-&\nearrow \atop \searrow&
   \left.\begin{array}{l}
   \frac{3}{2}^-\\
   \frac{5}{2}^-\\
   \end{array}\right\}\quad
   \{ \Lambda_{Qk2}^{**} \} \nonumber\\
 \left\{ q_1 q_2 \right\}:1^- &\nearrow \atop \searrow&
   \left.\begin{array}{l}
   \frac{1}{2}^-\\
   \frac{3}{2}^-\\
   \end{array}\right\}
   \quad \{ \Sigma_{Qk1}^{**} \}
 \eea
 \zeilee
\end{itemize}
One thus has altogether seven $\Lambda$-type p-wave states and seven
$\Sigma$-type $p$-wave states. The analysis can easily be extended to the
case $SU(6)\otimes 0(3)$ bringing in the strangeness quark in addition.

Let us mention that, in the charm sector the states $\Lambda_c(2285)$ and
$\Sigma_c(2453)$ are well established while there is first evidence for
the $\Sigma_c^*(2510)$ state. Recently two excited states $\Lambda_c^{**}
(2593)$ and $\Lambda_c^{**}(2627)$ have been seen which very likely
correspond to the two $p$-wave states making up the $\Lambda_{cK1}^{**}$
Heavy Quark Symmetry doublet. The charm-strangeness states $\Xi_c(2470)$
and $\Omega_c(2720)$ have been seen and first evidence was presented for
the $\frac{1}{2}^+ \; \Xi'_c(2570)$ state with the flavour configuration
$c\{sq\}$. In the bottom sector the $\Lambda_b(5640)$ has made its way
into the Particle Data Booklet listing while some indirect evidence has
been presented for the $\Xi_b(5800)$.

\section{Heavy Baryon Spin Wave Functions}
We are now in the position to write down the covariant spin wave functions
of the heavy baryons. These will be needed to derive the predictions of
Heavy Quark Symmetry for current, pion and photon transitions. One has a
light diquark system with $J^P$ quantum numbers $j^P$ which couples with a
heavy quark with $J^P=1/2^+$ to form a degenerate pair of states with
$J^P=(j\pm1/2)^P$. The coupling goes according to the scheme

\begin{equation}
\begin{tabular}{ccccc}
& & & &$(j-1/2)^P$\\[-.5cm]
& & &$\nearrow$\\
$j^P$&$\otimes$&$1/2^+$\\
& & &$\searrow$\\[-.5cm]
& & & &$(j+1/2)^P$\\
light diquark& &heavy quark& &heavy baryon\\
$\varphi^{\mu_1\cdots\mu_j}$& & & &$u^{(\mu_1)\mu_2\cdots\mu_j}$\\
\end{tabular}
\end{equation}

\noindent
where $\varphi^{\mu_1\cdots\mu_j}$ stands for the spin wave function of the
light diquark with spin $j$ and where $u^{\mu_1\mu_2\cdots\mu_j}$ and
$u^{\mu_2\cdots\mu_j}$ are the Rarita-Schwinger spinor wave functions of
the degenerate pair of heavy baryons with $J=j+1/2$ and $J=j-1/2$,
respectively. The case $j=0$ is special. In this case there is a coupling
only to one heavy baryon state with $J=1/2$ as e.g. for the
$\Lambda$-type baryon ground state $\Lambda_Q$.

It is then straightforward to write down the covariant spin wave functions
$\Psi_{\alpha\beta\gamma}$ of the Heavy Quark Symmetry baryon doublets.
One has

\begin{equation}
\Psi_{\alpha\beta\gamma}=(\varphi_{\mu_1\cdots\mu_j})_{\alpha\beta}
\left\{{M_{\gamma\gamma'}^{\mu_1}{u_{\gamma'}}^{\mu_2\cdots\mu_j}\atop
M_{\gamma\gamma'}{u_{\gamma'}}^{\mu_1\cdots\mu_j}}\right\}
\label{eqa}
\end{equation}
where we have explicitly written out the Dirac spinor indices $\alpha$,
$\beta$ and $\gamma$. The spinor indices $\alpha$ and $\beta$ refer to the
light quark system and the index $\gamma$ refers to the heavy quark.
Dropping the spinor indices one has

\beq
\Psi=\varphi_{\mu_1\cdots\mu_j}\psi^{\mu_1\cdots\mu_j}
\label{eqb}
\eeq
where the ``superfield'' heavy baryon wave function $\psi^{\mu_1\cdots\mu_j}$
stands for the two spin wave functions $\{j-1/2,j+1/2\}$ as indicated in
Eq.(\ref{eqa}). What remains to be done is to determine the matrices
$M^{\mu_1}$ and $M$ in Eq.(\ref{eqa}). They can be worked out by noting
that the heavy baryon spin wave function has to satisfy the mass-shell
condition on the heavy quark spinor index, i.e.
\beq
\not\!v\psi^{\mu_1\cdots\mu_j}=\psi^{\mu_1\cdots\mu_j}
\label{eqc}
\eeq
Eq.(\ref{eqc}) is solved by
\begin{eqnarray}
M^{\mu_1}&=&N_j\gamma_\perp^{\mu_1}\gamma_5\nonumber\\
M&=&N'_j\cdot \eins
\end{eqnarray}
where the transverse gamma matrix $\gamma_\perp^\mu$ is defined by
$\gamma_\perp^\mu=\gamma^\mu-\not\!vv^\mu$ and where $v^\mu$ is the
four-velocity of the heavy baryon $v^\mu=p^\mu/M$. The normalization of the
coupling matrices $M^\mu$ and $M$ is fixed by the normalization condition
\beq
\bar\psi^{\mu_1\cdots\mu_j}\psi_{\mu_1\cdots\mu_j}=(-1)^{J-1/2}2M
\eeq
which gives $N_0=0$, $N_1=\sqrt{1/3}$, $N_2=\sqrt{1/10}$ and $N'_0=0$,
$N'_1=N'_2=1$ for the $s$- and $p$-wave cases discussed in this review.
There is an implicit understanding that the set of tensor indices
``$\mu_1\cdots\mu_j$'' is always completely symmetrized, traceless with
regard to any pair of indices and transverse to the line of flight in every
index. This is annoted explicitly in Table~\ref{tab1} and~\ref{tab2} where
the $s$- and $p$-wave heavy baryon wave functions are listed. For example,
the notation  $\{\mu_1^\perp\mu_2^\perp\}_0$ implies symmetrization,
tracelessness and transversity of the two tensor indices $\mu_1\mu_2$ as
specified above. For the sake of completeness we have also included the
light-side spin wave functions as given in a constituent quark model
approach for the light side, where
$\hat\chi^0=\frac1{2\sqrt{2}}(\vslash +1)C$ and
$\hat\chi^{1\mu}=\frac1{2\sqrt{2}}(\vslash+1)\gamma^\mu_\perp C$
with $C$ the charge conjugation matrix.

\begin{table}
\bec
\tcaption{\label{tab1}Spin wave functions (s.w.f.) of $\Lambda$-type $s$-
      and $p$-wave heavy baryons. Light-side spin wave functions are
      constituent spin wave functions.}
\vspace{5mm}
\renewcommand{\baselinestretch}{1.2}
\small \normalsize
\begin{tabular}{*{5}c}
\hline \hline
& \begin{tabular}{c}
    light side s.w.f.\\
    ${\hat\phi}^{\mu_{1} \dots \mu_{j}}$
  \end{tabular}
& $j^{P}$
& \begin{tabular}{c}
    heavy side s.w.f.\\
    $\psi_{\mu_{1} \dots \mu_{j}}$
  \end{tabular}
&$J^{P}$
\\ \hline \hline
$\Lambda_{Q}$&$\hat \chi$&$ 0^{+}$&$u$&$\frac{1}{2}^{+}$\\
\hline
$\{ \Lambda_{QK1}^{**} \} $&$ {\hat \chi}^{0} K_{\perp}^{\mu_{1}}$&
$1^{-}$&
    $\begin{array}{r}
      \frac{1}{\sqrt{3}} \gamma^{\perp}_{\mu_{1}} \gamma_{5} u\\
      u_{\mu_{1}}
     \end{array} $
   &$\begin{array}{c}
     \frac{1}{2}^{-} \\
     \frac{3}{2}^{-}
     \end{array}$
\\ \hline
$\Lambda_{Qk0}^{**}$&${\frac{1}{\sqrt 3}\hat \chi}^{1} \cdot
k_{\perp}$&$0^{-}$&
 $u$&$\frac{1}{2}^{-}$
\\ \hline
$\{ \Lambda_{Qk1}^{**} \}$&$\frac{i}{\sqrt 2} \varepsilon (\mu_{1} {\hat
\chi}^{1} k_{\perp} v)$&$1^{-}$&
    $ \begin{array}{r}
      \frac{1}{\sqrt{3}} \gamma^{\perp}_{\mu_{1}}\gamma_{5}u\\
      u_{\mu_{1}}
      \end{array} $
  &$  \begin{array}{c}
      \frac{1}{2}^{-} \\
      \frac{3}{2}^{-}
      \end{array} $
\\ \hline
$\{ \Lambda_{Qk2}^{**} \} $&$ \frac{1}{2} \{ {\hat \chi}^{1,\mu_{1}}
k_{\perp}^{\mu_{2}} \}_{0}$&$2^{-}$&
    $\begin{array}{r}
     \frac{1}{\sqrt{10}} \gamma_{\{ \mu_{1}}^{\perp} \gamma_{5}
                               u_{\mu_{2} \}_{0} }^{\hspace{1mm}}\\
     u_{\mu_{1} \mu_{2} }
     \end{array} $
  &$\begin{array}{c}
    \frac{3}{2}^{-}\\
    \frac{5}{2}^{-}
    \end{array}$
\\ \hline \hline
\end{tabular}
\end{center}
\renewcommand{\baselinestretch}{1}
\small \normalsize
\vspace{1cm}
\tcaption{\label{tab2}Spin wave functions (s.w.f.) of $\Sigma$-type $s$-
          and $p$-wave heavy baryons.}
\vspace{5mm}
\renewcommand{\baselinestretch}{1.2}
\small \normalsize
\begin{center}
\begin{tabular}{*{5}c}
\hline \hline
& \begin{tabular}{c}
     light side s.w.f.\\
     ${\hat\phi}^{\mu_{1} \dots \mu_{j}} $
  \end{tabular}
& $j^{P}$
& \begin{tabular}{c}
     heavy side s.w.f.\\
     $\psi_{\mu_{1} \dots \mu_{j}}$
  \end{tabular}
& $J^{P}$
\\ \hline \hline
$\{ \Sigma_{Q} \}$ &${\hat \chi}^{1\mu_{1}}$& $1^{+}$&
        $\begin{array}{r}
             \frac{1}{\sqrt{3}}\gamma^{\perp}_{\mu_{1}} \gamma_{5}u\\
             u_{\mu_{1}}
         \end{array} $
       &$\begin{array}{c}
             \frac{1}{2}^{+}\\
             \frac{3}{2}^{+}
         \end{array} $
\\ \hline
$\{ \Sigma_{Qk1}^{**} \}$&${\hat \chi}^{0} k^{\mu_{1}}_{\perp}$&$1^{-}$&
       $\begin{array}{r}
            \frac{1}{\sqrt{3}}\gamma^{\perp}_{\mu_{1}} \gamma_{5} u\\
            u_{\mu_{1}}
        \end{array}$
     &$\begin{array}{c}
            \frac{1}{2}^{-}\\
            \frac{3}{2}^{-}
        \end{array}$
\\ \hline
$\{ \Sigma_{QK0}^{**}\} $&${\frac{1}{\sqrt 3}\hat\chi}^{1} \cdot K_{\perp}$&
$0^{-}$&$u$&$ \frac{1}{2}^{-}$
\\ \hline
$\{ \Sigma_{QK1}^{**}\}$&$ \frac{i}{\sqrt2} \varepsilon(\mu_{1} {\hat \chi}
^{1} K_{\perp}v)$&$1^{-}$&
       $\begin{array}{r}
            \frac{1}{\sqrt{3}} \gamma^{\perp}_{\mu_{1}} \gamma_{5} u\\
            u_{\mu_{1}}
       \end{array}$
     &$\begin{array}{c}
            \frac{1}{2}^{-} \\
            \frac{3}{2}^{-}
       \end{array}$
\\ \hline
$\{ \Sigma_{QK2}^{**} \}$&$ \frac{1}{2} \{ {\hat \chi}^{1,\mu_{1}} K_{
\perp}^{\mu_{2}} \}_{0}$&$2^{-}$&
     $\begin{array}{r}
           \frac{1}{\sqrt{10}}\gamma^{\perp}_{ \{ \mu_{1}} \gamma_{5}
                                     u_{\mu_{2} \}_{0} }^{\hspace{1mm}}\\

           u_{\mu_{1} \mu_{2}}
      \end{array}$
    &$\begin{array}{c}
          \frac{3}{2}^{-} \\
          \frac{5}{2}^{-}
      \end{array}$
\\ \hline \hline
\end{tabular}
\renewcommand{\baselinestretch}{1}
\small \normalsize
\end{center}
\end{table}
\vfill\eject
\section{Generic Picture of Current, Pion and Photon Transitions}
In Fig.~\ref{transitions} we have drawn the generic diagrams that describe
$b\rightarrow c$
current transitions, and $c\rightarrow c$ pion and photon transitions
between heavy baryons in the Heavy Quark Symmetry limit. The heavy-side and
light-side transitions occur completely independent of each other (they
``factorize'') except for the requirement that the heavy side and the light
side have the same velocity in the initial and final state, respectively,
which are also the velocities of the initial and final heavy baryons. The
$b\rightarrow c$ current transition induced by the flavour-spinor
matrix~$\Gamma$ is hard and accordingly there is a change of velocities
$v_1\rightarrow v_2$, whereas there is no velocity change in the pion and
photon transitions. The heavy-side transitions are completely specified
whereas the light-side transitions $j_1^{p_1}\rightarrow j_2^{p_2}$,
$j_1^{p_1}\rightarrow j_2^{p_2}+\pi$ and
$j_1^{p_1}\rightarrow j_2^{p_2}+\gamma$ are described by a number of form
factors or coupling factors which parametrize the light-side transitions.
The pion and the photon couple only to the light side. In the case of the
pion this is due to its flavour content. In the case of the photon the
coupling of the photon to the heavy side involves a spin flip which is down
by $1/m_Q$ and thus the photon couples only to the light side in the Heavy
Quark Symmetry limit.
\begin{figure}
\bec
\input{uebergaenge.pstex_t}
\fcaption{\label{transitions}Generic picture of bottom to charm current
transitions, and pion and photon transitions in the charm sector in the
Heavy Quark Symmetry limit $m_Q\rightarrow\infty$}
\eec
\end{figure}

Referring to Fig.~1 we are now in the position to write down the generic
expressions for the current, pion and photon transitions according to the
spin-flavour flow depicted in Fig. \ref{transitions}. One has

\vspace{1cm}
\noindent
current transitions:
\begin{equation}
\bar\psi_2^{\nu_1\cdots\nu_{j_2}}\Gamma\psi^{\mu_1\cdots\mu_{j_1}}
\left(\sum_{i=1}^Nf_i(\omega)
t^i_{\nu_1\cdots\nu_{j_2};\mu_1\cdots\mu_{j_1}}\right)
\label{eqd}
\end{equation}
\begin{eqnarray}
n_1\cdot n_2&=&1\qquad N=j_{\mini}+1\nonumber\\
n_1\cdot n_2&=&-1\quad N=j_{\mini}\nonumber
\end{eqnarray}

\noindent
pion transitions:
\begin{equation}
\bar\psi_2^{\nu_1\cdots\nu_{j_2}}\psi_1^{\mu_1\cdots\mu_{j_1}}
\left(\sum_{i=1}^Nf_i^\pi
t^i_{\nu_1\cdots\nu_{j_2};\mu_1\cdots\mu_{j_1}}\right)
\label{eqe}
\end{equation}
\begin{eqnarray}
n_1\cdot n_2&=&1\qquad N=j_{\mini}\nonumber\\
n_1\cdot n_2&=&-1\quad N=j_{\mini}+1\nonumber
\end{eqnarray}

\noindent
photon transitions:
\begin{equation}
\bar\psi_2^{\nu_1\cdots\nu_{j_2}}\psi^{\mu_1\cdots\mu_{j_1}}
\left(\sum_{i=1}^Nf_i^\gamma
t^i_{\nu_1\cdots\nu_{j_2};\mu_1\cdots\mu_{j_1}}\right)
\label{eqf}
\end{equation}
\begin{eqnarray}
j_1&=&j_2\qquad N=2j_1\nonumber\\
j_1&\neq&j_2\quad N=2j_{\mini}+1\nonumber
\end{eqnarray}
where the $\psi^{\mu_1\cdots\mu_j}$ are the heavy baryon spin wave
functions introduced in Sec.~3.

In each of the above cases we have also given the result of counting the
number~$N$ of independent form factors or coupling factors. These are easy
to count by using either helicity amplitude counting or $LS$ partial wave
amplitude counting. In the case of current and pion transitions the
counting involves the normalities of the light-side diquarks which is
defined by $n=(-1)^jP$.

All three coupling expressions can also be written down in terms of
Wigner's 6-$j$ symbols as will be discussed later for the pion and photon
transitions. The generic expressions Eq.(\ref{eqd}), Eq.(\ref{eqe}) and
Eq.(\ref{eqf}) completely determine the heavy quark symmetry structure of
the current, pion and photon transitions. What remains to be done is to
write down independent sets of covariant coupling tensors. This will be done
in the next section.

\section{Coupling Structure of Current, Pion and Photon Transitions}
What remains to be done is to tabulate explicit expressions for the tensors
\linebreak $t^i_{\nu_1\cdots\nu_{j_2};\mu_1\cdots\mu_{j_1}}$
describing the light-side
transitions. We shall treat the current, pion and photon transitions in
turn and shall work out some sample transitions in order to familiarize the
reader with the use of the tables.
\\[1cm]
{\em Current transitions:}
\\[0.5cm]
The tensors $t^i_{\nu_1\cdots\nu_{j_2};\mu_1\cdots\mu_{j_1}}$ have to be
build from the vectors $v_1^{\nu_i}$ and $v_2^{\mu_i}$, the metric tensors
$g_{\mu_i\mu_k}$ and, depending on parity, from the Levi-Civita object
$\varepsilon(\mu_i\nu_kv_1v_2):=\varepsilon_{\mu_i\nu_k\alpha\beta}
v_1^\alpha v_2^\beta$. The relevant tensors of interest have been listed in
Table~\ref{tab3}. For the $1^+\rightarrow 1^+$ and $1^+\rightarrow 2^-$
transitions the number of linearly independent tensors is $N=2$. We have
chosen to diagonalize the light-side transition in terms of partial-wave
amplitudes with definite partial wave $L_V$. Table~\ref{tab3} also contains
some normalization information in that all the amplitudes square up to
$(\omega^2-1)^{L_V}$.

As an application consider the current induced transitions
$\Lambda_b\rightarrow\Lambda_c,{\Lambda_c}^{**}$. They involve the
following diquark transitions:
\vspace{0.5cm}
\bec
\renewcommand{\baselinestretch}{1.3}
\small\normalsize
\begin{tabular}{rlcllll}
$\Lambda_b$&$\rightarrow\Lambda_c$&:&$0^+$&$\rightarrow 0^+$
&allowed & $(N=1)$\\
&$\rightarrow\{\Lambda_{cK1}^{**}\}$&:&&$\rightarrow 1^-$
&allowed&$(N=1)$\\
&$\rightarrow\Lambda_{ck0}^{**}$&:&&$\rightarrow 0^-$
&forbidden&\\
&$\rightarrow\{\Lambda_{ck1}^{**}\}$&:&&$\rightarrow 1^-$
&allowed &$(N=1)$\\
&$\rightarrow\{\Lambda_{ck1}^{**}\}$&:&&$\rightarrow 2^-$
&forbidden\\
\end{tabular}
\vspace{-0.9cm}
\beq\eeq
\eec
\vspace{0.5cm}
where the curly bracket notation indicates that the particle symbol stands
for a Heavy Quark Symmetry doublet. It is noteworthy that Heavy Quark
Symmetry predicts that only four of the possible seven transitions to the
$p$-wave states are allowed.
\begin{table}
\tcaption{\label{tab3}Tensor structure of diquark transitions. Partial wave
$L_V$ is the partial wave of the transition
$j_1^{P_1} \rightarrow j_2^{P_2} +0^+$. Sign of the product of normalities
$n_1 \cdot n_2$ determines the number $N$ of independent transitions or
Isgur-Wise functions.}
\vspace{5mm}
\renewcommand{\baselinestretch}{1.8}
\small \normalsize
\begin{center}
\begin{tabular}{rlccl}
\hline \hline
\multicolumn{2}{c}{diquark transition}&partial wave &\hspace{3cm}&
    covariant coupling\\
\multicolumn{2}{c}{$j_{1}^{P_{1}} \rightarrow j_{2}^{P_{2}} $}&
    $L_V$&$n_1 \cdot n_2$&
    $t^{i}_{\mu_{1} \dots \mu_{j_{1}};\nu_{1} \dots \nu_{j_{2}}}$\\
\hline \hline
$0^{+}\rightarrow $&$ 0^{+}$&$0$&$+1$&$1$\\
\hline
$0^{+} \rightarrow$&$ 0^{-}$&forbidden&$-1$&-\\
 &$ 1^{-}$&1&$+1$&$v_{1\mu}$\\

 &$ 2^{-}$&forbidden&$-1$&-\\
\hline
$1^{+} \rightarrow$&$ 1^{+}$&$0$&$+1$&$\frac{1}{\sqrt{3}} \left( g_{\mu\nu}
-\frac{1}{\omega +1} v_{1\mu} v_{2\nu} \right)$\\
&&$2$&$+1$&$\frac{1}{\sqrt{6}} \left( (\omega^2-1) g_{\mu\nu} - (\omega +2)
 v_{1\mu}v_{2\nu}  \right)$\\
\hline
$1^{+} \rightarrow$&$ 0^{-}$&$1$&$+1$&$v_{2\mu}$\\
 &$ 1^{-}$&$1$&$-1$&$\frac{1}{\sqrt{2}}\varepsilon(\mu_1\nu_1 v_1 v_2)$\\
 &$ 2^{-}$&$1$&$+1$&$\sqrt{\frac{3}{5}}\left( v_{1\mu_1} g_{\mu_2 \nu_1}
    -\frac{1}{\omega +1} v_{1\mu_1} v_{1\mu_2} v_{2\nu_1} \right)$\\
 &&$3$&$+1$&$\frac{1}{\sqrt{10}}\left(2(\omega^2-1) v_{1\mu_1} g_{\mu_2 \
 nu_1}\right.$\\
 &&&& $\left.\qquad  -(2\omega +3) v_{1\mu_1} v_{1\mu_2} v_{2\nu_1} \right)$\\
\hline \hline
\end{tabular}
\end{center}
\renewcommand{\baselinestretch}{1}
\small \normalsize
\end{table}
Using the relevant entries in Tables~\ref{tab1} and~\ref{tab3} one can then
write down the Heavy Quark Symmetry structure of the full transition
amplitudes. One has
\begin{itemize}
\item[i)]$\Lambda_b\rightarrow\Lambda_c$\qquad$1/2^+\rightarrow 1/2^+$
\begin{equation}
M^\lambda=\bar u_2\Gamma^\lambda u_1f^{(0)}(\omega)\qquad f^{(0)}(1)=1
\end{equation}
\item[ii)]$\Lambda_b\rightarrow\{\Lambda_{cK1}^{**}\}$\qquad
  $1/2^+\rightarrow\left\{{1/2^-\atop 3/2^-}\right\}$
\begin{equation}
M^\lambda=\left\{{-\frac1{\sqrt 3}\bar u_2\gamma_5{\gamma_\perp}_2^\mu
  \atop\bar u_2^\mu}\right\}\Gamma^\lambda u_1f_1^{(1)}(\omega)v_{1\mu}
\end{equation}
\item[iii)]$\Lambda_b\rightarrow\{\Lambda_{ck1}^{**}\}$\qquad
  $1/2^+\rightarrow\left\{{1/2^-\atop 3/2^-}\right\}$
  \begin{equation}\end{equation}
  the same as case ii) with $f_1^{(1)}(\omega)\rightarrow f_2^{(1)}(\omega)$
\end{itemize}
In the case of a left-chiral current transition as in the Standard Model
one has $\Gamma^\lambda=\gamma^\lambda(1-\gamma_5)$. For the elastic
transition $\Lambda_b\rightarrow\Lambda_c$ one has the normalization
condition $f^{(0)}(\omega=1)=1$ at equal velocities $v_1=v_2$ due to the
normalization of the diquark states, whereas there is no such normalization
condition for the transition form factors $f_1^{(1)}(\omega)$ and
$f_2^{(1)}(\omega)$ at zero recoil $\omega=1$.

One can then easily calculate the contributions of $\Lambda_c$,
$\Lambda_{cK1}^{**}$ and $\Lambda_{ck1}$ to the Bj\o rken sum
rule\cite{landg} by squaring the relevant transition amplitudes and, in
the case of $\Lambda_{cK1}^{**}$ and $\Lambda_{ck1}^{**}$, by summing over
the contributions of the two degenerate partners in the respective doublets.
One obtains

\begin{equation}
1=|f^{(0)}(\omega)|^2+(\omega^2-1)(|f_1^{(1)}(\omega)|^2
+|f_2^{(1)}(\omega)|^2+\ \ldots\ )
\end{equation}

It is noteworthy that the transition $\Lambda_b\rightarrow\Lambda_{ck1}^{**}$
is predicted to be zero when the light-side transition is calculated in a
constituent quark model approach with $SU(2N_f)\times O(3)$ symmetry. The
reason is that the light-side transition involves a $s=0$ to a $s=1$ quark
spin transition  which is zero in the constituent picture. This would then
imply that there are altogether only two nonzero transitions to the seven
$\Lambda_c^{**}$ $p$-wave states, which, in the light of the Bj\o rken sum
rule, would imply that the quasi-elastic transition
$\Lambda_b\rightarrow\Lambda_c$ constitutes a large fraction of the total
inclusive semileptonic $\Lambda_b\rightarrow X_c$ decay rate.
\\[1cm]
{\em Pion transitions:}
\\[0.5cm]
The $(j_1+j_2)$ rank tensors $t^i_{\nu_1\cdots\nu_{j_2};\mu_1\cdots\mu_{j_1}}$
describing the light-side transitions $j_1^{P_1}\rightarrow j_2^{P_2}+\pi$
have to be composed from the building blocks
$g_{\perp\mu\nu}=g_{\mu\nu}-v_\mu v_\nu$,
$p_{\perp\mu}=p_\mu-p\cdot v\,v_\mu$ and, depending on parity, from the
Levi-Civita tensor $\varepsilon(\mu_i\nu_kp\,v)$. In the case when there
are two independent transitions we have diagonalized the light-side
transition by going to the $LS$-basis. The tensors and amplitudes are now
labelled by the partial wave $l_\pi$ of the pion emission process. Again, we
have introduced some normalization information in Table~\ref{tab4}. The
normalization of the partial wave amplitudes $f_{l_\pi}$ is such that a
given partial wave amplitude $f_{l_\pi}$ contributes as
$|f_{l_\pi}|^2|\vec p|^{2l_\pi}$ to the spin-summed square of the diquark
transition amplitude.
\begin{table}
\tcaption{\label{tab4}Tensor structure of pion couplings to diquark states.
The pion is in a definite orbital state $l_{\pi}$. Tensor structure of
transitions with $(j_{1}^{P_{1}},j_{2}^{P_{2}})\rightarrow(j_{1}^{-P_{1}}
,j_{2}^{-P_{2}})\rightarrow(j_{2}^{P_{2}},j_{1}^{P_{1}})\rightarrow
(j_{2}^{-P_{2}},j_{1}^{-P_{1}})$ are identical and are not always listed
here.}
\vspace{5mm}
\renewcommand{\baselinestretch}{1.2}
\small \normalsize
\begin{center}
\begin{tabular}{rlcl}
\hline \hline
\multicolumn{2}{c}{diquark transition}& orbital wave & covariant coupling\\
\multicolumn{2}{c}{$j_{1}^{P_{1}} \rightarrow j_{2}^{P_{2}} + \pi$}&
    $l_{\pi}$& $t^{i}_{\mu_{1} \dots \mu_{j_{1}};\nu_{1} \dots \nu_{j_{2}}}$\\
\hline \hline
$0^{+}\rightarrow $&$ 0^{+}+\pi$&forbidden&-\\
\hline
$1^{+} \rightarrow$&$ 0^{+} +\pi$&1&$p_{\mu_{1}}^{\perp}$\\
 &$ 1^{+}+\pi$&1&$\frac{1}{\sqrt{2}}\varepsilon (\mu_{1} \nu_{1} p v )$\\
\hline
$0^{-} \rightarrow$&$0^{+}+\pi$&0&1 (scalar)\\
  &$ 1^{+}+\pi$&forbidden&-\\
  &$ 0^{-}+\pi$&forbidden&-\\
\hline
$1^{-} \rightarrow$&$ 0^{+}+\pi$&forbidden&-\\
  &$ 1^{+}+\pi$&$0$&$\frac{1}{\sqrt{3}} g_{\mu_{1} \nu_{1}}^{\perp}$\\
  &&$2$&$\sqrt{\frac{3}{2}}(p_{\mu_{1}}^{\perp} p_{\nu_{1}}^{\perp}-\frac{1}{3}
         p_{\perp}^{2}g_{\mu_{1} \nu_{1}}^{\perp})$\\
  &$ 0^{-}+\pi$&$1$&$ p_{\mu_{1}}^{\perp}$\\
  &$ 1^{-} +\pi$&$1$&$\frac{1}{\sqrt{2}} \varepsilon (\mu_{1} \nu_{1} pv)$\\
\hline
$2^{-} \rightarrow $&$ 0^{+}+\pi$&$2$&$\sqrt{\frac{3}{2}}p_{\mu_{1}}^{\perp}
                       p_{\mu_{2}}^{\perp}$\\
  &$ 1^{+}+\pi$&$2$&$p_{\mu_{2}}^{\perp} \varepsilon (\mu_{1} \nu_{1} p v )$\\
 &$ 0^{-}+\pi$&forbidden&-\\
 &$1^{-}+\pi$ &$1$&$\sqrt{\frac{3}{5}}g_{\mu_{1}\nu_{1}}^{\perp} p_{\mu_{2}}^
   {\perp}$\\
 &&$3$&$\sqrt{\frac{5}{2}} \{p_{\mu_{1}}^{\perp} p_{\mu_{2}}^{\perp} p_
      {\nu_{1}}^{\perp}-\frac{1}{5}(p_{\perp}^{2} g_{\mu_{1} \mu_{2}}^{\perp}
      p_{\nu_{3}}^{\perp}+$cycl.$(\mu_{1} \mu_{2} \nu_{1}))\}$\\
 &$ 2^{-}+\pi$ &$1$&$\sqrt{\frac{2}{5}}g_{\mu_{1}\nu_{1}}^{\perp}\varepsilon
      (\mu_{2} \nu_{2} p v )$\\
 &&$3$&$\sqrt{\frac{2}{5}}(p_{\mu_{1}}^{\perp}p_{\nu_{1}}^{\perp}-\frac{1}{5}
      g_{\mu_{1}\nu_{1}}p_{\perp}^{2})\varepsilon(\mu_{2} \nu_{2} p v)$\\
\hline \hline
\end{tabular}
\end{center}
\renewcommand{\baselinestretch}{1}
\small \normalsize
\end{table}
As an example we write down the pion transition amplitudes for the ground
state to ground state transition $\{\Sigma_c\}\rightarrow\Lambda_c+\pi$.
Using Tables~\ref{tab1} and~\ref{tab2} for the heavy-side baryon wave
functions and Table~\ref{tab4} for the $1^+\rightarrow 0^++\pi$ light-side
pion transition one has
\beq\label{sila-trans}
M^{\pi}={\bar u}_{2}(v) \left \{
   \begin{array}{c}
       \frac{1}{\sqrt{3}} \gamma_{\perp}^{\mu} \gamma_{5} u_{1}(v)\\
       u^{\mu}(v)
   \end{array}
\right \} f_{p} p_{\mu}^{\perp}
\eeq
Calculating the decay rate in the degeneracy limit $M_{\Sigma_c^*}=M_{
\Sigma_c}=M_1$ one finds
\beq\label{equal rates}
\Gamma_{\Sigma_{c}^{*} \rightarrow \Lambda_{c} + \pi}=\Gamma_{\Sigma_{c}
\rightarrow \Lambda_{c} + \pi}=\frac{1}{6\pi} \frac{M_{2}}{M_{1}} \mid
f_{p} \mid^{2} \mid \vec{p} \mid^{3}
\eeq
That the decay rates from degenerate doublet partners into a singlet state
are equal is a general result. This general result is much easier to derive
in the 6-$j$ symbol  approach than in the covariant approach
used so far. Looking again at the pion transition in Fig.~1 one sees that
one has to perform altogether three angular couplings. They are
\begin{itemize}
\item[i)]\hspace{1.7cm}${j_1}^{P_1}\otimes 1/2^+\Rightarrow {J_1}^{P_1}$
\item[ii)]\hspace{1.7cm}${j_2}^{P_2}\otimes 1/2^+\Rightarrow {J_2}^{P_2}$
\item[iii)]\hspace{1.7cm}${J_2}^{P_2}\otimes L_\pi\Rightarrow {J_1}^{P_1}$
\end{itemize}
\begin{equation}\end{equation}
where $L_\pi=l_\pi$ is the orbital momentum of the pion and ${J_1}^{P_1}$
and ${J_2}^{P_2}$ denote the $J^P$ quantum numbers of the initial and final
baryons. The heavy quark has $1/2^+$ quantum numbers. This is a coupling
problem well-known from atomic and nuclear physics and the problem is
solved by Wigner's 6-$j$ symbol calculus. One finds
\bea
M^{\pi}(J_{1}J_{1}^{z}\rightarrow J_{2}J_{2}^{z}+L_{\pi}m)
  &=&M_{L_{\pi}} (-1)^{L_{\pi}+j_{2}+1/2+J}(2j_{1}+1)^{1/2}(2J_{2}+1)^{1/2}
     \nonumber\\
  & &\hspace{0.5cm}\left \{ \begin{array}{ccc}
       j_{2}&j_{1}&L_{\pi}\\
       J    &J_{2}&1/2
     \end{array} \right \}
     \langle LmJ_{2} J_{2}^{z}\mid J_{1} J_{1}^{z} \rangle.
  \label{pi-trans3}
\eea
where
${\scriptscriptstyle \left \{  \begin{array}{ccc}
                          j_{2} & j & L_{\pi}\\
                          J_{1} & J_{2} & 1/2
                        \end{array} \right\}}$
is Wigner's 6-$j$ symbol and $\langle  L_\pi M J_2 J_2^z|J_1 J_1^z
\rangle$ is the Clebsch-Gordan coefficient coupling $L_\pi$ and $J_2$ to
$J_1$. $M_{L_\pi}$ is the reduced amplitude of the transition and is
proportional to $f_{l_\pi}$. Then by using the standard orthogonality
relation for the 6-$j$ symbols
one immediately concludes that the pion decay rates from degenerate doublet
partners into a singlet state (or vice versa) are equal.\cite{landh,landi}

\begin{figure}
\bec
\input{dubdub.pstex_t}
\vspace{0.4mm}
\fcaption{One-pion transition strengths for the transitions $\{
  \Lambda_{QK2}^{**}\} \rightarrow \{ \Sigma_Q\} + \pi$. Degeneracy levels are
  split for illustrative purposes.}
\eec
\end{figure}
Similarily one can calculate the doublet to doublet transition rates for
e.g. \linebreak $\{\Lambda_{Qk2}^{**}\}\rightarrow\{\Sigma_Q\}+\pi$.
The rates are in
the ratios $4:14:9:9$ as represented in Fig.~2. This result can easily be
calculated using the 6-$j$ formula Eq.(\ref{pi-trans3}) but involves
infinitely more labour in the covariant approach. Also, the result
``$4+14=9+9$'' for doublet to doublet one-pion transitions is a general
result which again can  easily be derived using the 6-$j$
approach.\cite{landh,landi}
\\[1cm]
{\em Photon transitions:}
\\[0.5cm]
In the photon transition case ${j_1}^{P_1}\rightarrow{j_2}^{P_2}+\gamma$
one has to use the field strength tensor
$F_{\alpha\beta}=k_\alpha\varepsilon_\beta-k_\beta\varepsilon_\alpha$ or,
depending on parity, its dual
$\tilde F_{\alpha\beta}=\frac12\varepsilon_{\alpha\beta\gamma\delta}
F^{\gamma\delta}$ in order to guarantee a gauge invariant coupling of the
photon to the light side. As in the pion transition case further building
blocks for the diquark transition tensor are the metric tensor, the
velocity $v_\alpha$ and the photon momentum $k_\mu$. The diquark photon
transitions listed in Table~\ref{tab5} are labelled by the total angular
momentum $J_\gamma$ of the photon (spin of the photon plus its orbital angular
momentum). The amplitudes are normalized such that the spin summed square
of a given diquark transition amplitude $f^{J_\gamma}$ is
$|f^{J_\gamma}|^2|\vec k|^{2J_\gamma+1}$.
\begin{table}
\tcaption{\label{tab5}Tensor structure of photon couplings to diquark states.
Photon is in definite multipole state EJ (electric) MJ (magnetic). Sign of
the product of naturalities determines whether coupling is to field
strength tensor $F_{\alpha \beta} \; (n_{1} \! \cdot \! n_{2}=+1)$ or to its
dual ${\tilde F}_{\alpha \beta} \; (n_{1} \! \cdot \! n_{2}=-1)$. Tensor
structure of transitions with $(j_{1}^{P_{1}}, j_{2}^{P_{2}}) \rightarrow
(j_{1}^{-P_{1}}, j_{2}^{-P_{2}}) \rightarrow (j_{2}^{P_{2}} j_{1}^{P_{1}})
\rightarrow (j_{2}^{-P_{2}} j_{1}^{-P_{1}})$ are identical and are not always
listed separately. }
\begin{center}
\vspace{5mm}
\renewcommand{\baselinestretch}{1.2}
\small \normalsize
\begin{tabular}{rlcccl}
\hline \hline
\multicolumn{2}{c}{\begin{tabular}{c}
                      diquark transition\\
                      $j_{1}^{P_{1}} \rightarrow j_{2}^{P_{2}}+\gamma$
                   \end{tabular}}
  &multipoles&$n_{1}n_{2}$&\hspace{0.5cm}&
                   \begin{tabular}{c}
                      covariant coupling\\
                      $t^{i}_{\mu_{1}\dots \mu_{j_{1}};\nu_{1}
                      \dots\nu_{j_{2}}}$
                   \end{tabular}\\
 \hline \hline
 $0^{+} \rightarrow$&$0^{+}+\gamma$&forbidden&$+1$&&\\
 \hline
 $1^{+} \rightarrow$&$0^{+}+\gamma$&M1&$-1$&$$&$\frac{1}{\sqrt{2}} {\tilde F}_
        {\alpha \beta}g_{\mu_{1}}^{\alpha}v^{\beta}$\\
     &$1^{+}+\gamma$&M1&$+1$&$$&$ \frac{1}{2}F_{\alpha \beta}g_{\mu_{1}}
        ^{\alpha}g_{\nu_{1}}^{\beta}$\\
     &&E2&$+1$&$$&$ \frac{1}{2}F_{\alpha
\beta}(2k_{\mu_{1}}g_{\nu_{1}}^{\alpha}
        v^{\beta}+k \hspace{-0.7mm}\cdot\hspace{-0.7mm} vg_{\mu_{1}}^{\alpha}
        g_{\nu_{1}}^{\beta})$\\
 \hline
 $0^{-} \rightarrow$&$0^{+}+\gamma$&forbidden&$-1$&$$&\\
 \hline
 $1^{-} \rightarrow $&$ 0^{+} + \gamma$&E1&$+1$&$$&$\frac{1}{\sqrt{2}}F_{\alpha
       \beta} g_{\mu_{1}}^{\alpha} v^{\beta}$\\
   &$1^{+} +\gamma$&E1&$-1$&$$&$\frac{1}{2}{\tilde F}_{\alpha \beta}
       g_{\mu_{1}}^{\alpha} g_{\nu_{1}}^{\beta}$\\
   &&M2&$-1$&$$&$\frac{1}{2}{\tilde F}_{\alpha \beta}(2k_{\mu_{1}}
       g_{\nu_{1}}^{\alpha}v^{\beta}+k \hspace{-0.7mm}\cdot\hspace{-0.7mm}v
       g_{\mu_{1}}^{\alpha} g_{\nu_{1}}^{\beta})$\\
 \hline
 $2^{-} \rightarrow$&$0^{+} +\gamma$&M2&$-1$&$$&${\tilde F}
       _{\alpha \beta} k_{\mu_{1}} g_{\mu_{2}}^{\alpha} v^{\beta}$\\
   &$1^{+} + \gamma$&E1&$+1$&$$&$ \sqrt{\frac{3}{10}}F_{\alpha \beta}
          g_{\mu_{1}}^{\alpha} g_{\mu_{2}\nu_{1}}v^{\beta}$\\
   &&M2&$+1$&$$&$\sqrt{\frac{1}{6}}F_{\alpha \beta} ( v \! \cdot \! k
          g_{\mu_{2} \nu_{1}} g_{\mu_{1}}^{\alpha}v^{\beta} + 2k_{\mu_{2}}
          g_{\mu_{1}}^{\alpha}g_{\nu_{1}}^{\beta} )$\\
   &&E3&$+1$&$$&$
        \sqrt{\frac{1}{30}}F_{\alpha \beta} ( (v \! \cdot \! k)^2  g_{\mu_{2}
          \nu_{1}}g_{\mu_{1}}^{\alpha} v^{\beta}$\\
   && &&&$\qquad + \frac{5}{4} v \! \cdot \! k
          k_{\mu_{2}} g_{\mu_{1}}^{\alpha} g_{\nu_{1}}^{\beta} $\\
   && &&&$\qquad + \frac{15}{4} v^{\beta} k_{\mu_{2}} ( k_{\nu_{1}}
          g_{\mu_{1}}^{\alpha} + k_{\mu_{1}} g_{\nu_{1}}^{\alpha} ))$\\
 \hline
 $2^{-} \rightarrow$&$0^{-}+\gamma$&E2&$+1$&$$&$F_{\alpha \beta}k_{\mu_{1}}
          g_{\mu_{2}}^{\alpha}v^{\beta}$\\
   &$1^{-}+\gamma$&M1&$-1$&$$&$\sqrt{\frac{3}{10}}{\tilde F}_{\alpha \beta}
          g_{\mu_{1}}^{\alpha}g_{\mu_{2} \nu_{1}} v^{\beta}$\\
   &&E2&$-1$&$$&$\sqrt{\frac{1}{6}}{\tilde F}_{\alpha \beta} ( v \! \cdot \! k
          g_{\mu_{2} \nu_{1}} g_{\mu_{1}}^{\alpha}v^{\beta} + 2 k_{\mu_{2}}
          g_{\mu_{1}}^{\alpha}g_{\nu_{1}}^{\beta} )$\\
   &&M3&$-1$&$$&$\sqrt{\frac{1}{30}}{\tilde F}_{\alpha \beta} ((v\!\cdot\!k)^2
          g_{\mu_{2} \nu_{1}} g_{\mu_{1}}^{\alpha} v^{\beta}$\\
   && &&&$\qquad + \frac{5}{4} v \!
          \cdot \! k k_{\mu_{2}} g_{\mu_{1}}^{\alpha} g_{\nu_{1}}^{\beta} $\\
   && &&&$\qquad + \frac{15}{4} v^{\beta} k_{\mu_{2}} ( k_{\nu_{1}}
          g_{\mu_{1}}^{\alpha} + k_{\mu_{1}} g_{\nu_{1}}^{\alpha} ))$\\
   &$2^{-}+\gamma$&M1&$+1$&$$&$\sqrt{\frac{1}{5}} F_{\alpha \beta} g_{\mu_{1}
          \nu_{1}}g_{\mu_{2}}^{\alpha}g_{\nu_{2}}^{\beta}$\\
   &&E2&$+1$&$$&$\sqrt{\frac{3}{7}}F_{\alpha \beta} g_{\mu_{1} \nu_{1}}
          (2 k_{\mu_{2}} v^{\beta}g_{\nu_{2}}^{\alpha} + v \! \cdot \! k
          g_{\mu_{2}}^{\alpha} g_{\nu_{2}}^{\beta})$\\
   &&M3&$+1$&$$&$\sqrt{\frac{3}{10}}F_{\alpha \beta} g_{\mu_{2}}^{\alpha}
          g_{\nu_{2}}^{\beta} ( (v \! \cdot \! k)^2  g_{\mu_{1} \nu_{1}} +
          \frac{5}{2} k_{\mu_{1}} k_{\nu_{1}})$\\
   &&E4&$+1$&$$&$\sqrt{\frac{1}{14}}F_{\alpha \beta} (2 k_{\mu_{2}} v^{\beta}
          g_{\nu_{2}}^{\alpha}$\\
   && &&&$\qquad\qquad + v \! \cdot \! k g_{\mu_{2}}^{\alpha}
          g_{\nu_{2}}^{\beta}) ( (v \! \cdot \! k)^2  g_{\mu_{1} \nu_{1}}$\\
   && &&&$\qquad + \frac{7}{2} k_{\mu_{1}} k_{\nu_{1}} )$\\
 \hline \hline
\end{tabular}
\end{center}
\renewcommand{\baselinestretch}{1}
\small \normalsize
\end{table}
It is then an easy matter to derive the Heavy Quark Symmetry structure of
photon transitions between heavy baryon states using Tables~\ref{tab1},
\ref{tab2} and~\ref{tab5}. As an illustration we write down the amplitude for
the ground state transition $\{\Sigma_Q\}\rightarrow\Lambda_Q+\gamma$. One has
\begin{equation}
M^{\gamma}={\bar u}_{2}
    \left \{ \begin{array}{c}
      \frac{1}{\sqrt{3}} \gamma_{\perp}^{\mu_{1}}\gamma_{5} u_{1}\\
      u_{1}^{\mu_{1}}
    \end{array} \right \}
\frac{1}{\sqrt{2}}f^{M1} {\tilde F}_{\alpha \beta} g_{\mu_{1}}^{\alpha}
v^{\beta}
\label{eqg}
\end{equation}
Using standard $\varepsilon_{\alpha\beta\gamma\delta}$-tensor identities
one obtains
\begin{eqnarray}
  \Sigma_{c} \rightarrow \Lambda_{c}+\gamma &:&\quad M^{\gamma}=i \frac{1}
  {\sqrt{6}}f^{M1} {\bar u}_{2} \kslash \eslash^{*} u_{1}\\
\Sigma_{c}^{*} \rightarrow \Lambda_{c}+\gamma &:& \quad M^{\gamma}=
  \frac{1}{\sqrt 2}f^{M1} {\bar u}_{2} \varepsilon(\mu_{1} v k\varepsilon^{*})
   u_{1}^{\mu_{1}} \label{eqh}
\end{eqnarray}
The transition (\ref{eqh}) can be checked to have the correct $M1$
coupling structure. In the degeneracy limit $M_{\Sigma_c^*}=M_{\Sigma_c}=M_1$
one finds the rate expressions
\beq
\Gamma_{\Sigma_{c} \rightarrow \Lambda_{c}+\gamma}=\Gamma_{\Sigma_{c} ^{*}
\rightarrow \Lambda_{c}+\gamma}=\frac{1}{6\pi} \mid \! f^{M1} \! \mid^{2}
\frac {M_{2}}{M_{1}} \mid \vec{k} \mid^{3}
\eeq
where $|\vec k|=(M_1^2-M_2^2)/2M_1$. The equality of the decay rates of
heavy quark symmetry partners into the ground state $\Lambda_c$ is again a
general result that can easily be derived in the 6-$j$ formalism as applied
to photon transitions. In order to derive the relevant decay formula in the
6-$j$ approach one again compounds the three angular momentum
couplings\cite{landh,landj}
\begin{itemize}
\item[i)]\hspace{1.7 cm}${j_1}^{P_1}\otimes 1/2^+\Rightarrow{J_1}^{P_1}$
\item[ii)]\hspace{1.7 cm}${j_2}^{P_2}\otimes 1/2^+\Rightarrow{J_2}^{P_2}$
\item[iii)]\hspace{1.7 cm}${J_2}^{P_2}\otimes J_\gamma\Rightarrow{J_1}^{P_1}$
\end{itemize}
\begin{equation}\end{equation}
where the notation is identical to the pion case treated before except for
the replacement $L_\pi\rightarrow J_\gamma$ and where $J_\gamma$ is the total
angular momentum of the photon. The heavy baryon photon transition
amplitude may then be written as
\begin{eqnarray}
M^{\gamma}(J_{1} J_{1}^{z} \rightarrow J_{2} J_{2}^{z}+J_{\gamma}m)
&=&M_{J_{\gamma}}(-1)^{J_{\gamma}+j_{2}+\frac{1}{2}+J_{1}} (2j_{1}+1)
   ^{\frac{1}{2}} (2J_{2}+1)^{\frac{1}{2}}\nonumber\\
& &\left \{ \begin{array}{ccc}
             J_{\gamma}&j_{2}&j_{1}\\
             \frac{1}{2}&J_{1}&J_{2}
           \end{array} \right \}
   \langle J_{\gamma}mJ_{2}J_{2}^{z}\mid J_{1}J_{1}^{z}\rangle\label{pho6j}
\end{eqnarray}
The reduced matrix elements $M_{J_\gamma}$ correspond to the multipole
amplitudes $f^{J_\gamma}$ as e.g. in Eq.(\ref{eqg}). Again, using orthogonality
relations for the 6-$j$ symbols, one can deduce that the one-photon rates
of doublet partners into singlet states are equal. Similarily there is a sum
rule for photon transitions between doublets as discussed for the pion
transitions.\cite{landh,landj}

\section{Summary and Conclusion}
We have provided a comprehensive set of formulas that allow one to work out
the predictions of Heavy Quark Symmetry for current, pion and photon
transitions in the baryon sector. We have chosen to present the material in
a form which emphasizes the similarities between the three different types
of transitions. The formulation is general and easily extends to
transitions involving higher orbital excitations. The coupling of the
various angular momentum involved in the transitions has been done using
conventional covariant techniques, and, in the case of pion and photon
transitions, also by 6-$j$ coupling methods. Although we have chosen to
express our results for the pion and photon transitions in terms of
transition amplitudes the formulas can easily be transcribed to the
language of chiral and gauge invariant\cite{landi,landj} effective
Lagrangians. In conclusion one may state that we are certainly looking
forward to analyze the forthcoming wealth of data on heavy baryon decays
to see how Heavy Quark Symmetry is at work.

\end{document}